\documentclass[preprint, superscriptaddress, showpacs,
aps]{revtex4}
\usepackage{multirow}
\usepackage{graphicx}
\begin{document}
\title{Effects of level inversion on the Coulomb displacement energies of the Ar-K isobaric analog state pairs}
\author{Zaijun Wang}
\email{ zjwang@tute.edu.cn} \affiliation{School of Science,
Tianjin University of Technology and Education, Tianjin 300222,
China}
\author{Xiaoyong Guo}
\affiliation{School of Science, Tianjin University of Science and
Technology, Tianjin 300457, China}
\author{Renli Xu} \affiliation{School of Information Technology, Nanjing University of
Chinese Medicine, Nanjing 210023, China}

\noindent
\begin{abstract}
The Coulomb displacement energies of the neutron-rich Ar-K
isobaric analog state pairs with mass number $A=35-47$ are
calculated within the relativistic mean field model and the
effects of level inversion on the Coulomb displacement energies of
the Ar-K isobaric analog state pairs are studied. The calculations
are carried out in two cases, with and without consideration of
the possible $2s_{1/2}$ and $1d_{3/2}$ proton level inversion of
the neutron-rich Ar isotopes. Results show that the $2s_{1/2}$ and
$1d_{3/2}$ level inversion of the neutron-rich Ar isotopes may
reduce the Coulomb displacement energy by 0.06$\sim$0.17 MeV for
the Ar-K isobaric analog state pairs. The results may provide a
reference for experimental investigations of nuclear level
inversion and a new test of the relativistic mean field model.


\end{abstract}

\maketitle

One of the long debated problem in nuclear physics is the
existence of the level inversion and the relevant nuclear
phenomena in exotic nuclei \cite{wim1,www1}. The problem of
nuclear level inversion has been studied with a variety of nuclear
models and much progress has been achieved
\cite{Ots1,Ots2,Ots3,Uts,Pov,Cau,wil46,sie67,Swi83,Dec03}.
However, there are still many problems remaining unresolved; for
instance, what are the real causes for the nuclear level
inversion? In a recently published paper \cite{wang14}, we studied
the possible proton 2$s_{1/2}$ and 1$d_{3/2}$ level inversion of
the neutron-rich Ar isotopes within the relativistic mean field
model
\cite{Ren95,Ren02,zhou03,jiang03,jiang05,jiang05_2,Sha93,Sug94}
and investigated the possibility to probe the level inversion
using elastic electron-nucleus scattering. In addition to electron
scattering, nuclear level inversion may also lead to other
physical effects. It may, for instance, lead to the change of the
Coulomb displacement energy \cite{Ant97,Suz98,Com83} since the
Coulomb displacement energies between isobaric analog states are
due to isospin non-conserving forces, such as the Coulomb
interaction between protons. Therefore, we would like to seek
another possible way to detect the nuclear level inversion by
investigating the effects of proton level inversion on the Coulomb
displacement energies of the Ar-K isobaric analog state pairs.

In the present paper, we report our calculations of Coulomb
displacement energy shifts brought about by the 2$s_{1/2}$ and
1$d_{3/2}$ proton level inversion of the neutron-rich Ar isotopes.
Suggested by professor W. Mittig, we performed calculations on the
Coulomb displacement energies of some neutron-rich Ar-K isobaric
analog state pairs with $A=35-47$ within the relativistic mean
field model with the NL-SH \cite{Sha93} and TM1/TM2 \cite{Sug94}
parameters. The calculations were carried out in two cases: with
and without consideration of the possible $2s_{1/2}$ and
$1d_{3/2}$ level inversion of the neutron-rich Ar isotopes. The
results show that the $2s_{1/2}$ and $1d_{3/2}$ level inversion
may lead to a noticeable reduction of the Coulomb displacement
energy, and this may provide a reference for experimental
investigations of nuclear level inversion. In addition, the
results may also provide a new test of the relativistic mean field
theory.

\section{Description of the Method}
Based on the formula
\begin{eqnarray}
 \Delta E_{C}=\frac{e}{2T}\int
 \rho_{exc}(\textbf{r})V_{core}(\textbf{r})d^{3}\textbf{r}
\end{eqnarray}
given in Ref. \cite{Com83}, where $V_{core}(\textbf{r})$ is the
Coulomb potential produced by the core protons and
$\rho_{exc}(\textbf{r})$ is the distribution of the $N-Z$ excess
neutrons, the Coulomb displacement energies between the isobaric
analog states and the respective nuclear ground states can be
evaluated. We first carried out calculations on some arbitrarily
chosen isobaric state pairs with mass number ranging from $A=27$
to $A=65$ to test the formula and the codes, and then performed
calculations on neutron-rich Ar-K isobaric state pairs with mass
number ranging from $A=35$ to $A=47$.

In the calculation, the Coulomb potential field
$V_{core}(\textbf{r})$ produced by the core protons and the
distribution $\rho_{exc}(\textbf{r})$ of the $N-Z$ excess neutrons
are evaluated using one of the paired nuclei with the smaller
proton number $Z_{<}$. For instance, for the Ar-K pair with
$A=46$, the Coulomb potential field $V_{core}(\textbf{r})$ is
generated by the 18 core protons of Ar, that is, the 16 protons in
$1s_{1/2}$, $2p_{1/2}$, $2p_{3/2}$, $1d_{5/2}$, $2s_{1/2}$
orbitals and the 2 protons in the $1d_{3/2}$ orbital; the excess
neutron distribution $\rho_{exc}(\textbf{r})$ is given by the
$N-Z=10$ neutrons with 2 of them in $1d_{3/2}$ (or $2s_{1/2}$)
orbital and 8 in $1f_{7/2}$ and higher orbitals. The Coulomb
potential field $V_{core}(\textbf{r})$ and the excess neutron
distribution $\rho_{exc}(\textbf{r})$ are both calculated using
the single particle wave functions produced with the relativistic
mean field model. For comparison and mutual verification, two sets
of force parameters, the NL-SH \cite{Sha93} and TM1/TM2 parameters
\cite{Sug94}, are used, with TM1 used for the isobaric analog
state pairs with $A>40$ and TM2 for those with $A\leq40$.

\section{Numerical results and discussions}
\subsection{The arbitrarily chosen isobaric analog state pairs with $A=27-65$}
To guarantee the validity of our calculation, we first test the
method and the codes. We have arbitrarily chosen some isobaric
analog state pairs for the test. In Table 1, we tabulate the
calculated results and the experimental ones from Ref.
\cite{Ant97}, as well as the deviations along with some
statistical results. The first and second columns of Table 1 are
the mass numbers and corresponding specific isobaric analog state
pairs that are chosen and calculated. The mass number ranges from
$A=27$ to $A=65$. The numbers in the brackets of the second column
are the isospin values \cite{Ant97}. The symbol $\bigtriangleup
E_{C}$ in the third column denotes the experimental Coulomb
displacement energies \cite{Ant97}; the symbols $\bigtriangleup
E_{C}^{1}$ in
\begin{table}[!hbp]
\scriptsize \caption{The first column is the mass numbers of the
isobaric pairs. The third column is the experimental Coulomb
displacement energies \cite{Ant97}, and the fourth and fifth
columns are the calculated results with the NL-SH and TM1/TM2
parameters ( with TM1 parameters for the pairs with $A>40$ and TM2
parameters for the pairs with $A\leq40$ ). Columns six to nine are
the deviations between the calculated results and the experimental
ones. The last column is the renewed experimental data. The last
three rows are the mean, minimum and maximum deviations.}
\begin{tabular*}{16.3cm}{@{\extracolsep{\fill}}clcccccccl}\hline\hline
\multirow{2}{*}{A}&isobaric pair&$\bigtriangleup
E_{C}$&$\bigtriangleup E_{C}^{1}$&$\bigtriangleup
E_{C}^{2}$&\multirow{2}{*}{$|\bigtriangleup
E_{C}^{1}-\bigtriangleup E_{C}|$}&\multirow{2}{*}{
$\frac{|\bigtriangleup E_{C}^{1}-\bigtriangleup
E_{C}|}{\bigtriangleup E_{C}}$\%}&\multirow{2}{*}{$|\bigtriangleup
E_{C}^{2}-\bigtriangleup E_{C}|$}&\multirow{2}{*}{
$\frac{|\bigtriangleup E_{C}^{2}-\bigtriangleup E_{C}|}{\bigtriangleup E_{C}}$\%}&\multirow{2}{*}{$\bigtriangleup E_{C}^{\texttt{new}}$}\\
\cline{2-2}\cline{3-3}\cline{4-4}\cline{5-5}
 &$Z_{<}$-$Z_{>} (T)$ & Expt.&NL-SH&TM1(2)&&&&& \\
\hline
27&Al-Si(1/2)  &5.595  &5.63751& 5.52230&0.04251&0.75979&0.0727&1.29937&5.5947    \\
29&Si-P(1/2)   &5.725  &5.69597& 5.51713 &0.02903&0.50707&0.20787&3.63092&5.72498    \\
31&P-S(1/2)    &6.178  &6.05676& 5.87302 &0.12124&1.96245&0.30498&4.93655&6.18038    \\
34&S-Cl(1)     &6.274  &6.42437& 6.26309 &0.15037&2.39672&0.01091&0.17389&6.27395           \\
35&Cl-Ar(1/2)  &6.748  &6.76497& 6.59962 &0.01697&0.25148&0.14838&2.19887&6.74845            \\
37&Ar-K(1/2)   &6.931  &7.09705& 6.92081  &0.16605&2.39576&0.01019&0.14702&6.9298             \\
41&Ca-Sc(1/2)  &7.278  &7.20970& 7.17352&0.0683&0.93844&0.10448  &1.43556&7.27783\\
42&Ca-Sc(1)    &7.208  &7.19530& 7.15578&0.0127&0.17619&0.05222  &0.72447&7.20846\\
43&Sc-Ti(1/2)  &7.650  &7.53283& 7.48842&0.11717&1.53163&0.16158 &2.11216&7.64927\\
45&Ti-V(1/2)   &7.915  &7.85444& 7.80318&0.06056&0.76513&0.11182 &1.41276&7.91325\\
46&Ti-V(1)     &7.834  &7.84021& 7.78598&0.00621&0.07927&0.04802&0.61297&7.83477\\
47&V-Cr(1/2)   &8.234  &8.17386& 8.11559&0.06014&0.73039&0.11841 &1.43806&8.22721\\
50&Mn-Fe(1)    &8.934  &8.82008& 8.75047&0.11392&1.27513&0.18353&2.05429&8.91945\\
51&Mn-Fe(1/2)  &8.802  &8.80570& 8.73400 &0.0037&0.04204&0.068&0.77255&8.82376\\
53&Fe-Co(1/2)  &9.085  &9.10302& 9.02239&0.01802&0.19835&0.06261 &0.68916&9.07042\\
54&Co-Ni(1)    &9.578  &9.44369& 9.36090&0.13431&1.40228&0.2171  &2.26665&9.56907\\
55&Co-Ni(1/2)  &9.476  &9.40562& 9.34438&0.07038&0.74272&0.13162 &1.38898&9.47639\\
56&Mn-Fe(3)    &8.596  &8.55811& 8.59486&0.03789&0.44079&0.00114 &0.01326&8.59577\\
59&Cu-Zn(1/2)  &9.874  &9.73325& 9.73302&0.14075&1.42546&0.14098 &1.42779&9.92516\\
65&Ni-Cu(9/2)  &9.221  &9.21356& 9.19391&0.00744&0.08069&0.02709&0.29379&9.21991\\
\hline
mean&&&&&0.06888&0.91&0.10918&1.46&\\
min&&&&&0.0037&0.043&0.00114&0.014&\\
max&&&&&0.16605&2.40&0.30498&4.94&\\ \hline \hline
\end{tabular*}
\end{table}
the fourth column and $\bigtriangleup E_{C}^{2}$ in the fifth
column denote the calculated results of the Coulomb displacement
energies with the NL-SH and TM1/TM2 parameters, respectively.
Columns six and seven are the absolute and relative deviations of
the calculated values with NL-SH parameters with respect to the
experimental values. Columns eight and nine are the same as
columns six and seven but for the results calculated with TM1/TM2
parameters. In Table 1, the experimental data for the
$^{56}$Mn-$^{56}$Fe and $^{65}$Ni-$^{65}$Cu pairs were obtained
with both $^{56}$Fe and $^{65}$Cu in the excited states
\cite{Ant97}; for the other isobaric state pairs, the experimental
data were obtained with the paired nuclei both in the ground
states \cite{Ant97}. The last three rows in Table 1 are the mean,
minimum and maximum deviations, respectively. For intuitive
comparison, the results are also shown in Fig. 1.

\begin{figure}[!htb]
\includegraphics[width=11cm]{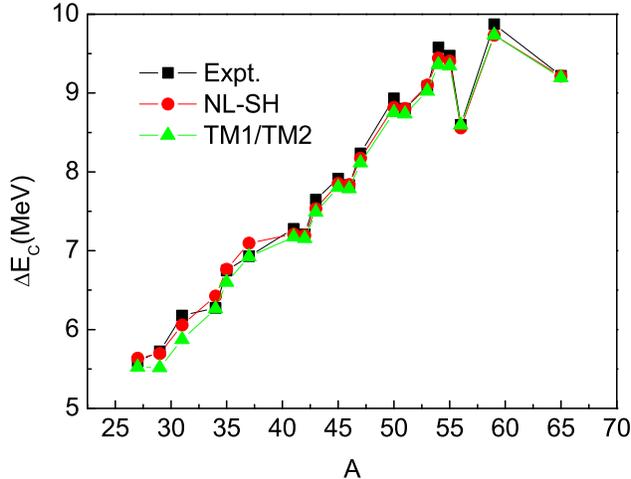}
\vspace{-1.6cm}\caption{Comparison of the calculated results with
the experimental data \cite{Ant97}. The filled squares denote the
experimental data; the filled dots, the calculated results with
the NL-SH parameters; the filled triangles, the results with the
TM1/TM2 parameters.}
\end{figure}

\begin{figure}
\includegraphics[width=11cm]{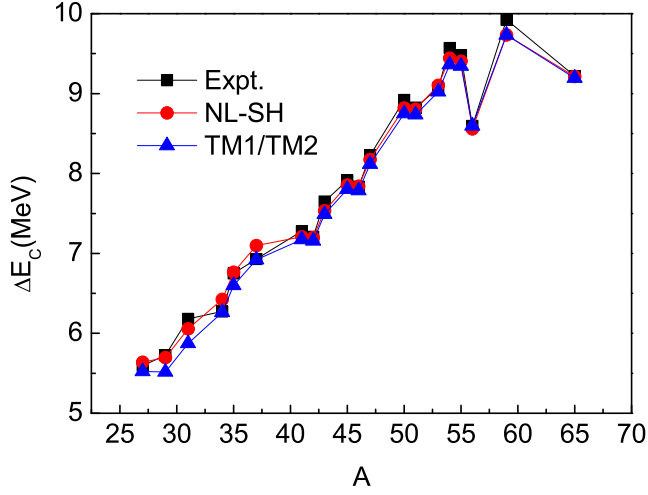}
\vspace{-1.8cm}\caption{Comparison of the calculated results with
the renewed experimental data. The filled squares denote the
renewed experimental data; the filled dots, the calculated results
with the NL-SH parameters; the filled triangles, the results with
the TM1(2) parameters.}
\end{figure}

The experimental results of Ref. \cite{Ant97} were obtained based
on a relatively old atomic nuclear mass data table given by Audi
and Wapstra in 1995 \cite{Aud95}. To compare the present results
with up-to-date experimental data, we renewed the experimental
results based on the most recent atomic nuclear mass data given by
Audi and Wapstra in 2012 \cite{Aud12}. The renewed experimental
data are evaluated with the same formula \cite{Ant97},
\begin{eqnarray}
 \Delta E_{C}=M_{Z>}-M_{Z<}+\Delta \texttt{nH},
\end{eqnarray}
as is used in Ref. \cite{Ant97}, where $M_{Z>}$ is the mass of the
higher $Z$ member nucleus of an analog pair and $M_{Z<}$ is the
mass of the lower $Z$ member nucleus, and $\Delta \texttt{nH}$ is
the neutron-hydrogen mass difference of $0.782354$ MeV. The
renewed data are listed in the last column of Table 1 and the
comparison with the calculated results is shown in Fig. 2.

Table 1, Fig. 1 and Fig. 2 show that the calculated Coulomb
displacement energies are in good agreement with the experimental
data. It is seen from Table 1 that the largest deviation of the
results obtained with NL-SH parameters is less than $0.17$ MeV and
the mean deviation is less than $0.069$ MeV; the corresponding
relative deviations are $2.40\%$ and $0.91\%$, respectively. For
the results obtained with TM1/TM2 parameters, the largest absolute
deviation is less than $0.31$ MeV and the mean deviation is $0.11$
MeV; the corresponding relative deviations are $4.94\%$ and
$1.46\%$. For the renewed data, it can also be easily obtained
from Table 1 that the largest deviation of the results obtained
with NL-SH parameters is less than $0.20$ MeV and the mean
deviation is less than $0.072$ MeV; the corresponding relative
deviations are $2.42\%$ and $0.94\%$, respectively. For the
results obtained with TM1/TM2 parameters, the largest absolute
deviation is less than $0.31$ MeV and the mean deviation is $0.12$
MeV; the corresponding relative deviations are $4.98\%$ and
$1.47\%$. The experimental data are very well reproduced by the
relativistic mean field model with the NL-SH and TM1/TM2
parameters. It is also found from Table 1 that the difference is
very small (no more than 0.06 MeV) between the renewed data and
the data of Ref. \cite{Ant97}.

\subsection{The Ar-K isobaric analog state pairs}
The Ar-K isobaric analog state pairs, which may be of some
interests to the experimentalists, are calculated and discussed.
The mass number of the chosen isobaric analog pairs ranges from
$A=35$ to $A=47$.

Calculations of the Ar isotopes within the relativistic mean field
model show that there may exist the proton $2s_{1/2}$ and
$1d_{3/2}$ level inversion for the Ar isotopes with $A>36$
\cite{wang14}. Since the $2s_{1/2}$ and $1d_{3/2}$ level inversion
in Ar isotopes will change the distribution of the $N-Z$ excess
neutrons, we calculated the Coulomb displacement energies of the
Ar-K isobaric analog state pairs in two cases. In one case (Case
1), the level inversion is neglected and two of the $N-Z$ excess
neutrons are assumed to occupy the $1d_{3/2}$ orbital and the
others occupy the $1f_{7/2}$ and higher level orbitals; in another
case (Case 2), the level inversion is taken into account and two
of the $N-Z$ excess neutrons are assumed to be in the $2s_{1/2}$
orbital and the others in the $1f_{7/2}$ and higher level
orbitals. As has been shown that the renewed data only show very
slight difference from the experimental results given in Ref.
\cite{Ant97}, so we will not give the renewed data in the
following calculations for the Ar-K isobaric analog state pairs.

\subsubsection{Case 1}
In this case, the $2s_{1/2}$ and $1d_{3/2}$ level inversion for
the Ar isotopes with $A>36$ is not considered, i.e., the
distribution $\rho_{exc}(\textbf{r})$ of the $N-Z$ excess neutrons
are evaluated with two of the $N-Z$ excess neutrons in the
$1d_{3/2}$ orbital and the others in the $1f_{7/2}$ and higher
level orbitals. The results are listed in Table 2.

In Table 2, the experimental data are from Ref. \cite{Ant97}; the
experimental result of $^{37}$Ar-$^{37}$K was obtained with
$^{37}$Ar and $^{37}$K both in the ground states; the others were
obtained with either Ar or K nucleus in the excited states, or
both of them in the excited states. The last row in Table 2 is the
results of the $^{47}$Ar-$^{47}$K pair, whose experimental data
are unavailable. The results are also plotted in Fig. 3, with
panel (a) plotted on the same scale as Fig. 1 for comparison and
panel (b) on a smaller scale to show some details for discussion
and comparison. The plots show that the general trends of
variation of the calculated results with the two sets of
parameters are very similar to that of the experimental data.

\begin{table}\scriptsize
\caption{The experimental Coulomb displacement energies of the
Ar-K isobaric pairs and the calculated results in the case that
the $2s_{1/2}$ and $1d_{3/2}$ level inversion of Ar isotopes is
not considered.}
\begin{tabular*}{16cm}{@{\extracolsep{\fill}}clccc}\hline \hline
\multirow{2}{*}{A}&isobaric pair&$\bigtriangleup
E_{C}$&$\bigtriangleup E_{C}^{1}$&$\bigtriangleup
E_{C}^{2}$\\
\cline{2-2}\cline{3-3}\cline{4-4}\cline{5-5} &$Z_{<}$-$Z_{>}(T)$ &
Expt.&NL-SH&TM1(2) \\
\hline 35&Ar-K(3/2)&  7.091 &  7.14234& 6.96404\\
36&Ar-K(1)& 6.977 &  7.12519& 6.96379\\
37&Ar-K(1/2)& 6.931& 7.09885 &6.92676\\
38&Ar-K(1)& 6.826& 7.051& 6.87161\\
39&Ar-K(3/2)& 6.764& 6.94675& 6.77747\\
40&Ar-K(2)& 6.671& 6.87715& 6.71587\\
41&Ar-K(5/2)& 6.64& 6.82366 &6.66754\\
45&Ar-K(9/2)& 6.515& 6.66499& 6.59396\\
46&Ar-K(5)& 6.553& 6.61714& 6.54256\\
47&Ar-K& --& 6.53535 &6.47139\\ \hline \hline
\end{tabular*}
\end{table}

\begin{figure}
\includegraphics[width=10cm]{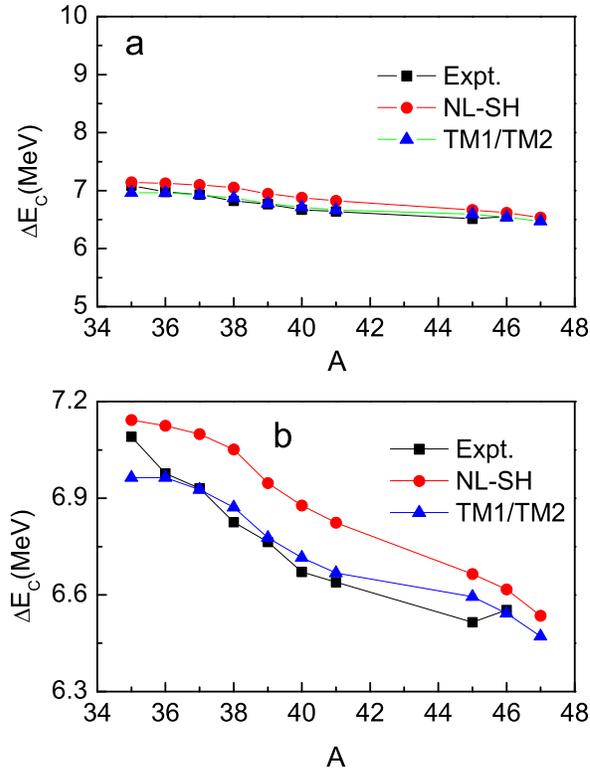}
\vspace{-3.0cm}\caption{The experimental data and calculated
results for the Ar-K isobaric analog state pairs without
consideration of the possible 2$s_{1/2}$ and 1$d_{3/2}$ state
level inversion of Ar isotopes. In panel (a) the results are
plotted on the same scale as Fig. 1 and in panel (b) the results
are plotted on a smaller scale.}
\end{figure}

However, on a smaller scale it is seen from panel (b) that in this
case the TM1/TM2 parameters produce better results for Ar-K pairs
than the NL-SH parameters, especially for those near the stability
line. The results obtained with the NL-SH parameters are generally
a little larger than the experimental values.

\subsubsection{Case 2}
In this case, the $2s_{1/2}$ and $1d_{3/2}$ level inversion for
the Ar isotopes with $A>36$ is taken into account in the
calculation of the Coulomb displacement energies. The distribution
$\rho_{exc}(\textbf{r})$ of the $N-Z$ excess neutrons are
evaluated with two of the $N-Z$ excess neutrons in the $2s_{1/2}$
orbital and the others in the $1f_{7/2}$ and higher level
orbitals. The numerical results are given in Table 3 and plotted
in Fig. 4 on the same scales as Fig. 3.
\begin{table}[!hbp]
\scriptsize \caption{The experimental results of the Ar-K isobaric
pairs and the calculated results in the case that the $2s_{1/2}$
and $1d_{3/2}$ level inversion of Ar isotopes is considered.}
\begin{tabular*}{16cm}{@{\extracolsep{\fill}}clccc}\hline \hline
 \multirow{2}{*}{A}&isobaric
pair&$\bigtriangleup E_{C}$&$\bigtriangleup
E_{C}^{1}$&$\bigtriangleup
E_{C}^{2}$\\
\cline{2-2}\cline{3-3}\cline{4-4}\cline{5-5} &$Z_{<}$-$Z_{>}(T)$ &
Expt.&NL-SH&TM1(2) \\
\hline 35&Ar-K(3/2)&  7.091&   7.14234& 6.96404\\
36&Ar-K(1)& 6.977 &  7.12519& 6.96379\\
37&Ar-K(1/2)& 6.931& 7.03958& 6.79593\\
38&Ar-K(1)& 6.826& 6.98583& 6.75248\\
39&Ar-K(3/2)& 6.764& 6.83119& 6.62398\\
40&Ar-K(2)&  6.671& 6.74552& 6.55085\\
41&Ar-K(5/2)& 6.64& 6.68764& 6.50013\\
45&Ar-K(9/2)& 6.515& 6.53733& 6.47287\\
46&Ar-K(5)& 6.553& 6.49264& 6.42513\\
47&Ar-K& --& 6.41374& 6.35766
\\ \hline \hline
\end{tabular*}
\end{table}

It is seen from Fig. 4 that, in this case, the results are lowered
or shifted down a little in comparison with the results of Case 1.
The experimental results nearly fall between the theoretical
results obtained from the two parameter sets; the deviations
between the results obtained with the NL-SH parameters and the
experimental ones are reduced, yet the results given by the NL-SH
parameters are still a little larger than the experimental
results; however, the deviations of the results given by the
TM1/TM2 parameters from the experimental ones are enlarged
slightly.

\begin{figure}[!htb]
\includegraphics[width=12cm]{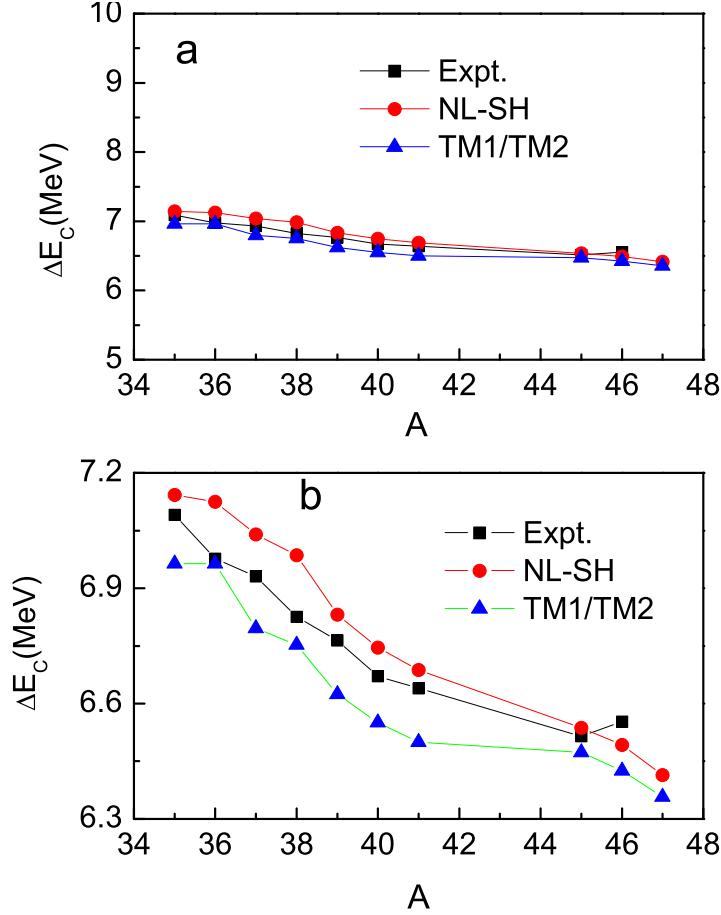}
\vspace{-2.9cm}\caption{The same as Fig.3 but for the results
obtained considering the possible 2$s_{1/2}$ and 1$d_{3/2}$ state
level inversion of the Ar isotopes.}
\end{figure}

To show the effects of the 2$s_{1/2}$ and 1$d_{3/2}$ state level
inversion on the Coulomb displacement energy, we calculated the
differences of the results between Case 1 and Case 2 for either
parameter set. The results are shown in Table 4 and plotted in
Fig. 5, where
\begin{eqnarray}
 \delta=\bigtriangleup
E_{C}^{1}(case 1)-\bigtriangleup E_{C}^{1}(case 2)
\end{eqnarray}
for the NL-SH parameters, and
\begin{eqnarray}
 \delta=\bigtriangleup E_{C}^{2}(case
1)-\bigtriangleup E_{C}^{2}(case 2)
\end{eqnarray}
for the TM1/TM2 parameters.

The results in Table 4 show that the $2s_{1/2}$ and $1d_{3/2}$
level inversion of Ar isotopes leads a reduction to the Coulomb
displacement energies by 0.06$\sim$0.17 MeV for the Ar-K isobaric
pairs, and the largest reduction, which occurs to the Ar-K pairs
near $A=41$, is about 0.17 MeV. A circumstantial evidence for our
result was given in Ref. \cite{Suz98}, where the effects of the
halo on the Coulomb displacement energy for the isobaric analog
state of $^{11}$Li were investigated, and the halo is found to
reduce the Coulomb displacement energy by 0.100$\sim$0.200 MeV.
The plots in Fig. 5 show that the trends of variation of the
differences $\delta$ given by the two sets of parameters are
approximately the same.

\begin{table}
\scriptsize \caption{The differences of the calculated Coulomb
displacement energies between the two cases for the two sets of
parameters.}
\begin{tabular*}{16cm}{@{\extracolsep{\fill}}clcc}\hline \hline
 \multirow{2}{*}{A}&isobaric
pair&$\delta$&$\delta$\\
\cline{2-2}\cline{3-3}\cline{4-4} &$Z_{<}$-$Z_{>}(T)$&NL-SH&TM1(2) \\
\hline
35&Ar-K(3/2)&   0& 0\\
36&Ar-K(1)&  0& 0\\
37&Ar-K(1/2)& 0.05927& 0.13083\\
38&Ar-K(1)& 0.06517& 0.11913\\
39&Ar-K(3/2)& 0.11556& 0.15349\\
40&Ar-K(2)& 0.13163& 0.16502\\
41&Ar-K(5/2)& 0.13602& 0.16741\\
45&Ar-K(9/2)& 0.12766& 0.12109\\
46&Ar-K(5)& 0.1245& 0.11743\\
47&Ar-K& 0.12161 & 0.11373\\ \hline \hline
\end{tabular*}
\end{table}

\begin{figure}[!htb]
\includegraphics[width=12cm]{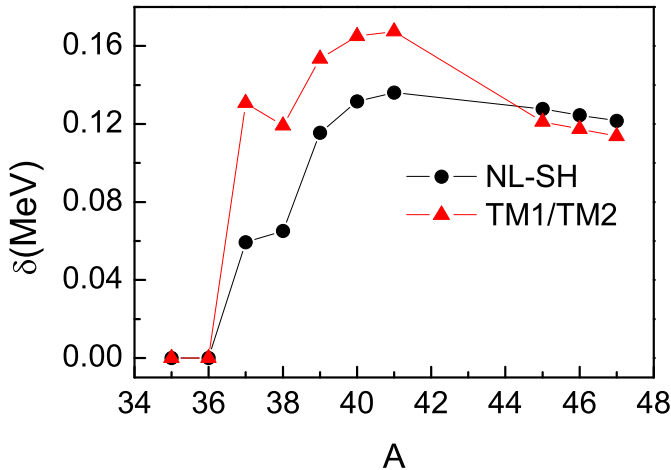}
\vspace{-2.2cm}\caption{Differences of the calculated results of
the Ar-K isobaric analog state pairs between Case 1 and Case 2 for
the two parameter sets.}
\end{figure}

Finally, a few remarks on the result of the $^{46}$Ar-$^{46}$K
pair should be made. It can be found from panel (b) of Fig. 3 or
Fig. 4 that the experimental result of the $^{46}$Ar-$^{46}$K pair
seems to have a ``strange behavior''; it shows a deviation, an up
warp, from the gradually decreasing trend of variation of the
Coulomb displacement energies of the Ar-K pairs with the increase
of mass number. This deviation is not reproduced by either set of
parameters in either cases. This may imply either that the
description of very neutron-rich Ar-K pair by the relativistic
mean field model is not so good as that of the nuclei near the
stability line or that there is a lack of accuracy in the
experimental result for the $^{46}$Ar-$^{46}$K pair, since the
excitation energy for $^{46}$K given in Ref. \cite{Ant97} is just
an approximate value ($\approx11.470$ MeV).

\section{Summary}
Within the relativistic mean field model, the Coulomb displacement
energies of the Ar-K isobaric analog state pairs with mass number
$A$ ranging from 35 to 47 are calculated directly with the Coulomb
potential generated by the core protons and the distribution of
the $N-Z$ excess neutrons within the relativistic mean field
model. The calculation is carried out in two cases: In one case
(Case 1), the 2$s_{1/2}$ and 1$d_{3/2}$ state level inversion is
not considered, i.e., two of the $N-Z$ excess neutrons are assumed
to be in the $1d_{3/2}$ orbital; in another case (Case 2), the
2$s_{1/2}$ and 1$d_{3/2}$ state level inversion is taken into
account, i.e., two of the $N-Z$ excess neutrons are assumed to be
in the $2s_{1/2}$ orbital. The results show that the 2$s_{1/2}$
and 1$d_{3/2}$ state level inversion may lead a reduction of no
more than 0.17 MeV to the Coulomb displacement energy, and the
largest reduction may occur to the $^{40-42}$Ar-$^{40-42}$K pair.
The results may provide a reference for experimentally studying
the 2$s_{1/2}$ and 1$d_{3/2}$ state level inversion. In addition,
the $^{46}$Ar-$^{46}$K pair is also discussed for its ``seemingly
strange'' experimental value.

\begin{center}
{\large Acknowledgments }
\end{center}
This work is Supported by the National Natural Science Foundation
of China (Grant No.11275138, No.11535004, No.11375086,
No.11120101005, No.10675090, and No.11404241), and by the Research
Fund of Tianjin University of Technology and Education (Grant No.
KJYB11-3).

\end{document}